\documentclass[]{aa}

\usepackage{txfonts}
\usepackage{graphicx}
\usepackage{color}
\usepackage{natbib}
\usepackage{rotating}
\usepackage{url}
\usepackage{amssymb}
\bibpunct{(}{)}{;}{a}{}{,} 

\makeatletter
\def\artanh{\mathop{\operator@font artanh}\nolimits}
\makeatother

\begin{document}

\title{Dark matter distribution in the merging cluster Abell 2163
  \thanks{Based on observations obtained with MegaPrime/MegaCam, a joint
  project of Canada-France-Hawaii Telescope (CFHT) and CEA/DAPNIA, at the
  Canada-France-Hawaii Telescope (CFHT) which is operated by the National
  Research Council (NRC) of Canada, the Institut National des Sciences
  de l'Univers of the Centre National de la Recherche Scientifique (CNRS)
  of France and the University of Hawaii.}  }
\author{G. Soucail\inst{1,2}
}

\offprints{G. Soucail, gsoucail@irap.omp.eu}

\institute{
  Universit\'e de Toulouse; UPS-Observatoire Midi-Pyr\'en\'ees; IRAP; Toulouse, France 
\and 
  CNRS; Institut de Recherche en Astrophysique et Plan\'etologie, 
  14 Avenue Edouard Belin, F--31400 Toulouse,  France
}

\date{Received Oct 17, 2011; accepted Jan 24, 2012} 

\abstract
{The cluster Abell~2163 is a merging system of several subclusters
with complex dynamics. It presents exceptional X-rays properties
(high temperature and luminosity), suggesting that it is a very
massive cluster. Recent 2D analysis of the gas distribution has
revealed a complex and multiphase structure.}
{This paper presents a wide-field weak lensing study of the dark
matter distribution in the cluster in order to provide an alternative
vision of the merging status of the cluster. The 2D mass
distribution is built and compared to the galaxies and gas
distributions.}
{ A Bayesian method, implemented in the \textsc{Im2shape} software, was
used to fit the shape parameters of the faint background galaxies and
to correct for PSF smearing. A careful color selection on the background
galaxies was applied to retrieve the weak lensing signal.  Shear signal was
measured out to more than 2~Mpc ($\simeq 12\arcmin$  from the center). The
radial shear profile was fit with different parametric mass profiles.
The 2D mass map is built from the shear distribution and used to identify
the different mass components. }
{The 2D mass map agrees with the galaxy distribution, while the total mass
inferred from weak lensing shows a strong discrepancy to the X-ray deduced
mass. Regardless of the method used, the virial mass $M_{200}$ falls
in the range 8 to 14 $10^{14} \ h_{70}^{-1} M_\odot$ inside the virial radius 
($\sim 2.0 \ h_{70}^{-1}$ Mpc), a
value that is two times less than the mass deduced from X-rays. The
central mass clump appears bimodal in the dark matter distribution,
with a mass ratio $\sim$ 3:1 between the two components. 
The infalling clump A2163-B is detected in weak lensing as an
independent entity. All these results are interpreted in the context
of a multiple merger seen less than 1Gyr after the main crossover.}
{}


\maketitle

\section{Introduction}
\label{sec:introduction}
Clusters of galaxies are powerful probes for cosmology because they
belong to the high-mass end of the halo mass function of collapsed
structures \citep{voit05}. The mass distribution within these structures
is representative of the nonlinear development of the accretion processes
during the cosmological evolution \citep{press74,frenk90}. But it is
still a challenge to securely relate observational quantities like
the X-ray luminosity and temperature, or the optical number counts and
velocity dispersion, to constitutive properties such as the gravitational
potential or the mass distribution \citep{reiprich02, vikhlinin09,
carlberg96}. Several internal processes also play major roles throughout
cosmological times. They can distort the simple scaling relations between
physical quantities initially set by the hierarchical clustering process
of structures formation.

The rich cluster of galaxies Abell 2163 is an interesting cluster that
displays several paradoxical properties. First, it is one of the richest
Abell clusters, at a redshift $z=0.201$, and among the most luminous
ones in X-rays ($L_X= 6.0 \times 10^{45} \ h_{50}^{-2} \ \mathrm{erg \
s}^{-1}$ in the 2--10 keV band, \cite{elbaz95}). Extensive analysis of
the global properties of this exceptional cluster have soon been conducted
after the initial detection of a very high X-ray temperature ($T_X =
14.6 \pm 0.9$ keV, measured with GINGA by \citet{arnaud92}). They have
rapidly pointed evidence of a nonisothermal gas distribution with a strong
temperature drop in the outskirts \citep{markevitch94, markevitch96b} and
inhomogeneities of the gas temperature in the center \citep{markevitch01}.
More recently, \citet{feretti01} have detected a very extended and powerful
radio halo, which they interpret as the tracer of a very hot nonthermal phase
in the ICM. Moreover, areas with a flatter radio spectral index and
higher energetic particles are found to be coincident with the most
disturbed regions of the X-ray distribution \citep{feretti04}. Another
component was identified as a secondary X-ray source in the map,
spectroscopically confirmed to be much cooler than the main cluster
(A2163-B). The strong temperature variations in the center displayed
by Chandra imaging \citep{govoni04} are also representative of ongoing
merging processes; however, the X-ray properties in the external regions
of the cluster are more regular, both in luminosity and temperature,
and fall within the general category of massive clusters.

Recently, \citet{maurogordato08} have presented a very detailed optical
analysis of this cluster that sheds new light on its merging status. Using
both spectroscopy of a large sample of galaxy members and wide field
multicolor imaging, they studied the galaxy distribution and confirm
that the main cluster is dynamically separated from A2163-B, although
they belong to the same complex. The galaxy density distribution of the
main cluster A2163-A is strongly elongated in the EW direction and shows a
bimodal distribution that depends on the luminosity range of the galaxies
tracing the distribution. For the faintest ones, the distribution is
split into two components, A1 and A2, for which the former is centered
on the brightest cluster galaxy BCG1 while the latter is slightly
shifted compared to the galaxy BCG2 (following the naming defined by
\citet{maurogordato08}). In addition, a strong velocity gradient in
the main clump is related to the spatial galaxy distribution and is
elongated in the NE/SW direction. All these properties point towards a
post-merging phase where the main component has just undergone a recent
merger along the elongation direction, nearly coplanar with the plane
of the sky, while A2163-B is infalling onto A2163-A. Finally, a very
detailed analysis of the complex X-ray emission, based on high-resolution
imaging with Chandra, coupled with spectro-imaging with XMM-Newton, has
brought new light on the complex gas distribution \citep{bourdin11}. The
authors clearly identify a cold gas clump in the direction of A2 but not
coincident with it. This clump may be seen crossing the main cluster's
hot-gas component, a scenario with features resembling those detected
in the so-called ‘‘Bullet cluster'' 1E 0657--56 \citep{markevitch02}.

To explore the physical state of the cluster and to better understand
the relationship between the distributions of the galaxies, the hot
ICM gas, and the dark matter, we propose in this paper to use the weak
gravitational lensing effect, which is directly related to the dark
matter content and its distribution. A2163 has already been observed
in weak lensing by several groups but without strong and conclusive
results. \citet{squires97} used a large and shallow image of the cluster
and found a modest value for the velocity dispersion of about 700 km
s$^{-1}$ (adjusted with a singular isothermal potential), although the
authors claim that the uncertainties are so great close to the center
that they could accept higher values up to 1000 km s$^{-1}$.  With deeper
data from the VLT but in a smaller field of view, \citet{cypriano04}
measured $\sigma_V = 1020 \pm 150$ km s$^{-1}$. Again this is a low
value compared to what could be expected from such a hot cluster and
to what is measured dynamically from the galaxies.  More recently,
\citet{radovich08} have claimed that they solve this discrepancy with
their new measure of the weak shear profile based on deep and wide
field images of the cluster. However, their fit is not conclusive,
so we decided to provide a new analysis of the weak lensing signal
specifically dedicated to the spatial distribution of the dark matter,
using the same CFHT/MegaCam data. We took special care in selecting
the background galaxies and remove the cluster contamination close to
the center as much as possible. Moreover, our goal is to compare the 2D
mass map with the X-ray map in order to characterize the link between
the dark matter and the baryonic mass better in this merging cluster.
Similar approaches have led to the spectacular results in the ``Bullet
cluster'' 1E0657--56 \citep{clowe06} in which the separation between
the collisional baryonic matter and the collisionless dark matter
is obvious and allows accurate identification of the merging stage
\citep{bradac06}. In parallel to this work, \citet{okabe11} used Subaru
weak lensing data to provide a similar 2D analysis. Fortunately, their
results are globally similar to ours, although some discrepancies remain,
and these are investigated in the paper.

The paper is organized as follow: in section 2 we describe the data
and the weak lensing analysis, insisting on the process selecting
the background galaxies. In section 3, an analysis of the central
arclets is proposed and some strong lensing properties are derived
for the cluster center. Section 4 presents the 2D weak
lensing reconstruction and the spatial distribution of the dark
matter in the field of the cluster, while in section 5 we discuss our
mass estimates. Finally we present some conclusions in section 6.
Throughout the paper we use $H_0 = 70\,\mathrm{km\,s^{-1}\,Mpc^{-1}}$,
$\Omega_\mathrm{M}=0.3$, $\Omega_\Lambda=0.7$. At the redshift of the
cluster ($z=0.2$), $1\arcsec$ corresponds to $3.3\,\mathrm{kpc}$ and
$1\arcmin$ to $200\,\mathrm{kpc}$. Magnitudes are given in the AB system.

\section{Image data and weak lensing analysis}
\label{sec:data}
\subsection{Data reduction and photometry}
\label{ssec:datared}
Imaging data were obtained at the Canada-France-Hawaii Telescope
with MegaCam, during the run 05AC12 (PI:
H. Hoekstra) and were retrieved from the CFHT archives at
CADC\footnote{This research used the facilities of the Canadian
Astronomy Data Centre operated by the National Research Council of
Canada with the support of the Canadian Space Agency.}. 
These data were already preprocessed by the CFHT Elixir
pipeline \citep{magnier04} and correspond to Master detrend images. 
Nineteen images of 600 seconds each were found: 15 in 
$r'$ and 4 in $g'$. The total integration time is 
2.5 hours in $r'$ and 40 minutes in $g'$.

Astrometric solutions for each CCD and then for each image were
computed using  SCAMP \citep{bertin06}, a tool developed at
Terapix\footnote{\url{http://terapix.iap.fr/}}. Absolute coordinates
of the stars detected in the field were selected from the USNO-B1
catalog. Internal accuracy for the calibration of the whole set of
images reached 0.06\arcsec , while the external accuracy was on the order
of 0.27\arcsec\ or equivalently 1.5 pixels. All the images were 
combined and geometrically rescaled using SWARP, with a fixed pixel size
of 0.186\arcsec. Final images in $g'$ and $r'$ cover one square degree.

\begin{figure}
  \centering
  \includegraphics[width=0.48\textwidth]{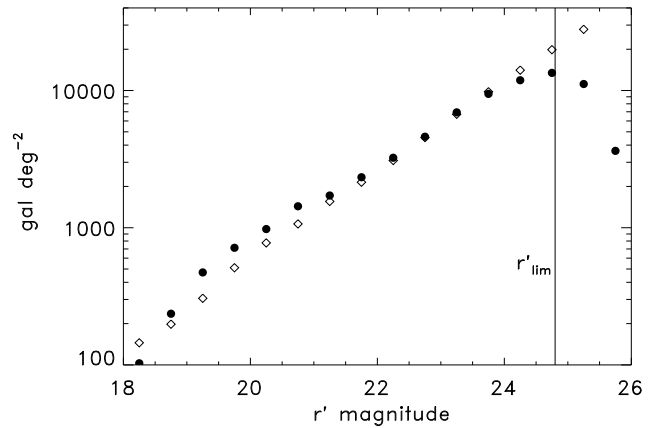}
  \caption{Galaxy number counts in the field of
Abell 2163. The black dots correspond to $r'$ magnitudes. The diamonds
are the $r'$ number counts in the CFHTLS deep field D1. In both fields,
only galaxies with CLASS\_STAR $<0.8$ are kept. The excess at bright
magnitudes in A2163 represents the cluster galaxies. The limiting magnitude
is consistent with what is expected for compact objects measured at the
$5\sigma$ level.}
  \label{fig:counts}
\end{figure}

Photometric internal calibration is provided during the observing run
and photometric zero points are measured and included in each image
header. But some images were not taken in photometric conditions. 
They were post-calibrated with reference to other available data, 
cross-correlating our catalogs with
those provided by \citet{maurogordato08}: these were obtained
with the ESO Wide Field Imager on the 2.2m telescope and the Bessel
R and V filters, under photometric conditions. In addition, they were
corrected from the galaxy extinction, with an average value $E(B-V) =
0.41$. Therefore, the magnitude difference between both sets of data,
taking the AB correction of the MegaCam filters into account, defines
the correction to add to the magnitude zero points: $-0.08$ mag in $r'$
and $-0.30$ mag in $g'$. Both corrections include the galactic extinction
correction and are estimated with an accuracy of 0.05 mag.

Photometric catalogs were built in both filters, using \textsc{SExtractor}
\citep{bertin96}. The $r'$ image, which is the deepest one, was
the reference image for detection, and the $g'-r'$ color index
was computed inside the limiting isophote defined in the $r'$
image for each object.  Completeness magnitudes were estimated
from number counts to $r'_{lim} = 24.8$ and $g'_{lim} = 25.3$
(Figure~\ref{fig:counts}).  They are approximately 0.7 brighter than
the $5\sigma$ limiting magnitudes computed inside an aperture with a
diameter 1.45 times the full width at half maximum (FWHM) of the seeing
disk, and defined in the MegaCam exposure time calculator (DIET\footnote{
\url{http://www.cfht.hawaii.edu/Instruments/Imaging/MegaPrime/dietmegacam.html}}).
This last magnitude is best suited to point-like objects, and that the
0.7 mag difference reflects that most of the faint objects are compact
but extended ones.  In the final images used in this paper, the measured
average seeing is 0.75\arcsec\ in $r'$ and 0.67\arcsec\ in $g'$.

\subsection{PSF correction and galaxy shape measurements}
\label{ssec:catalogs}
\begin{figure}
  \centering
  \includegraphics[width=0.48\textwidth]{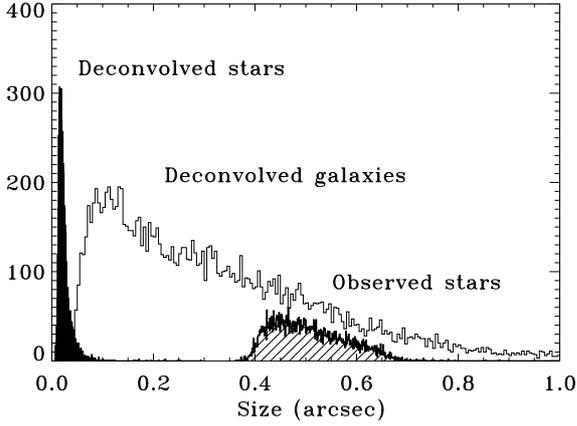}
  \caption{Histogram of the average size ($\sqrt{ab}$) of the
selected stars before (hatched histogram) and after (filled histogram)
the deconvolution process implemented with \textsc{Im2shape}.  For each
object, the local PSF is determined locally by the average of the 5
closest stars. }
  \label{fig:size_stars}
\end{figure}

Galaxies are first separated from stars with the help of the
magnitude-peak surface-brightness diagram, which clearly identifies
the stellar objects \citep{bardeau05}. Then for each galaxy, the local
PSF is measured by averaging the shape parameters of the five closest
stars. Intrinsic galaxy shape parameters are finally recovered using the
\textsc{Im2shape} software developed by \citet{bridle01}. This software
performs an analytical deconvolution of the galaxy images, approximated by
a single elliptical Gaussian, and returns the intrinsic shape parameters
with their uncertainties, estimated by a Bayesian analysis of the image
residuals. It has been successfully applied on a sample of clusters
of galaxies by \citet{bardeau07}, who developed a methodology for the
shear analysis that we follow fairly closely.  To test the validity of
the analysis chain, we checked that the result of the deconvolution
process applied to the stellar field itself shrinks the size of the
deconvolved stars to less than 0.1\arcsec, with a flat distribution of
their orientation (Fig.~\ref{fig:size_stars}). For the galaxies, the
histogram extends to a larger size, up to several arcseconds.  The whole
pipeline was also applied on the simulated sheared images developed
by the STEP consortium \citep{heymans06}. The results are presented in
detail in \citet{foex11}. In short, using simulated sheared images of
faint galaxies, the average measured shear is compared to the true shear
input in the simulations. This is done on a sequence of simulations with
increasing shear value and several sets of realistic PSFs.  The average
difference bewteen the shears is fit with a linear function of the true
shear. With our pipeline we systematically underestimate the true shear by
10\%, but with no systematic offset in the measured shear. Consequently,
we corrected the measures of the shear by applying a factor 1.1 to the
value of average tangential ellipticity of the galaxies. This correction
affects the total mass determination by $\sim 15\%$ upward.

\subsection{Selection of background galaxies}
\label{ssec:background_galaxies}
\begin{figure}
  \centering
  \includegraphics[width=0.48\textwidth]{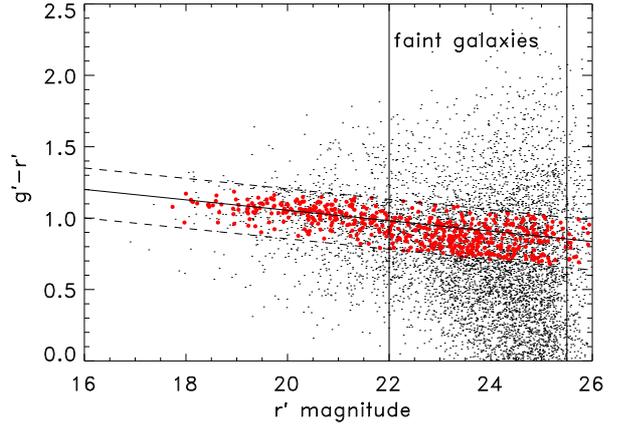}
  \caption{Color-magnitude plot of the galaxies in the field of Abell
2163, located within a 10\arcmin\ radius from the central galaxy BCG1.
The straight line corresponds to the cluster red sequence (RS), and the
dashed lines define the window for selecting ``cluster members''.
Galaxies located less than 200\arcsec\ from the cluster center and
within the selection window are plotted as large dots. }
  \label{fig:colormag}
\end{figure}

An important step in building a photometric catalog of objects for
weak lensing is to clean it from cluster and foreground contamination
as much as possible. The cluster contamination is the major source
of errors in the mass reconstruction for low redshift clusters: it
is most prominent in the center so it distorts the shear profile and
attenuates its central value, leading to an underestimation of the
central mass density.  In practice, we first selected the galaxies
within a magnitude range large enough to have a high number of objects
but limited in the bright end to remove a large fraction of cluster
members. The lower limit of the magnitude cut was chosen as $r'=22$, a
magnitude fainter than that of a $0.1 L^\star $ galaxy at the cluster
redshift. The upper limit was fixed to $r'=25.5$, {\it i.e.} 0.7
magnitude fainter than the completeness limit. This limit corresponds to
the $5 \sigma$ limiting magnitude of point sources estimated with DIET.
We added a selection criterion based on the color index.  This kind of
selection by the galaxies color properties has already been tested in
many studies \citep{kneib03,broadhurst05b,limousin07}. It has proved
to be quite powerful to remove galaxies with photometric properties
compatible with cluster members. We built the color-magnitude diagram
from the photometric catalog and identified the cluster red sequence
(Fig.~\ref{fig:colormag}). The average
color index of the bright ellipticals is $g'-r' = 1.15 \pm 0.05$ mag,
in good agreement with expected values from galaxy evolution codes
\citep{bruzual03}. All galaxies located $+0.15$ and $-0.20$ from
the red sequence line were stored in a catalog that we identify as the
``cluster galaxy'' catalog. Although it is not fully representative of the
color diversity of cluster members, it includes early-type galaxies that
dominate the density distribution of the galaxies. Moreover, a fraction
of bluer ones is also included because the lower color limit is slightly
widened. The galaxies outside this window and within the magnitude
range $[22.0-25.5]$ are stored for the weak lensing analysis. The
resulting catalog contains 14550 galaxies within a square area of side
37\arcmin\ centered on the galaxy BCG1. This sample is expected to
be substantially free of cluster contamination and is dominated by
background sources. With this selection, the mean density of sources
reduces from 18 to 12 galaxies/arcmin$^{2}$ in the central area of
the field, with a residual slight excess compared to the most external
regions of the field (no more than 10\%). Thus this corresponds to an
efficient cleaning of the cluster contamination.

\begin{figure}
  \centering
  \includegraphics[width=0.48\textwidth]{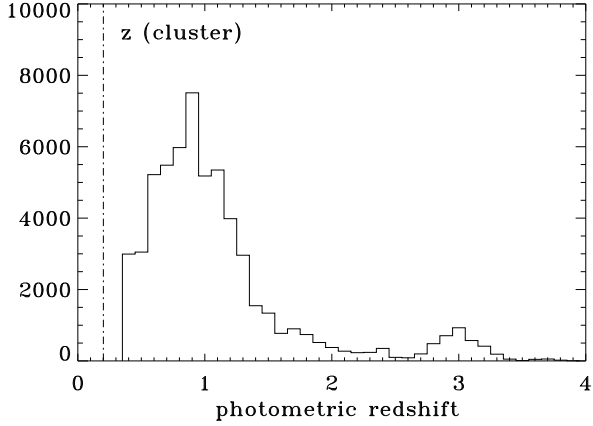}
  \caption{Photometric redshift distribution of the ``reference
catalog'' built from the CFHTLS-D1 data. The photometric selection applied
to the weak lensing catalog has been included (see text for details) as
well as a photometric redshift cut with $z_{\mathrm{zphot}} \geq 0.35$.}
  \label{fig:zphotdist}
\end{figure}

The quantitative mass reconstruction based on the measure of the
shear signal also requires an absolute scaling by the lensing factor
$<D_{LS}/D_{OS}>$, which depends on the redshift distribution of
lensed galaxies. In this ratio, $D_{LS}$ is the angular diameter
distance between the lens and the source while $D_{OS}$ is between
the source and the observer \citep{schneider92}. We do not have direct
information on this redshift distribution in the present case, so we
attempted to estimate it carefully. To do so, we built a photometric
catalog considered as a reference catalog, using the T0004 release of
the CFHTLS-Deep survey. Its main advantage is that the data have been
collected with the same instrument and the same filter set. But they
are much deeper than the present data (at least one magnitude deeper
in $g'$ and $r'$) and multicolor observations in five filters allow the
determination of accurate photometric redshifts. In practice we used the
photometric redshifts that have been publicly available after the T0004
release. They were carefully calibrated and validated with spectroscopic
samples \citep{coupon09}.  We applied to this catalog the same selection
criteria as we applied to the present catalog (magnitude cut $22<r'<25.5$
and color selection with $g'-r'$ outside the red sequence window), and
we only selected those objects having a photometric redshift $1\sigma$
error smaller than 0.15. We assumed that this sample is representative
of the galaxy population on the line of sight of A2163. Looking at
its redshift distribution, we measured a foreground contamination of
4\%. Moreover, for all the galaxies with a photometric redshift $z_{phot}
> z_{cluster} + 0.15$, we computed the average geometric factor for weak
lensing (Fig~\ref{fig:zphotdist}): \[ <\frac{D_{LS}}{D_{OS}}> = 0.72 \pm
0.10 \] with an average redshift for the sources $<z_{\mathrm{phot}}>
= 1.1$.  Our weak lensing catalog includes objects fainter than the
completeness limit ($r' \sim 24.8$) so a significant fraction of the
galaxies is probably missing in the range $[25-25.5]$. But we checked
that it does not significantly affect the redshift distribution, and the
average factor $<D_{LS}/D_{OS}>$ is not changed by more than a few \% .

\section{A two-dimensional mass map}
\label{sec:2d-mass-maps}
\begin{figure*}
  \centering
  \includegraphics[width=\textwidth]{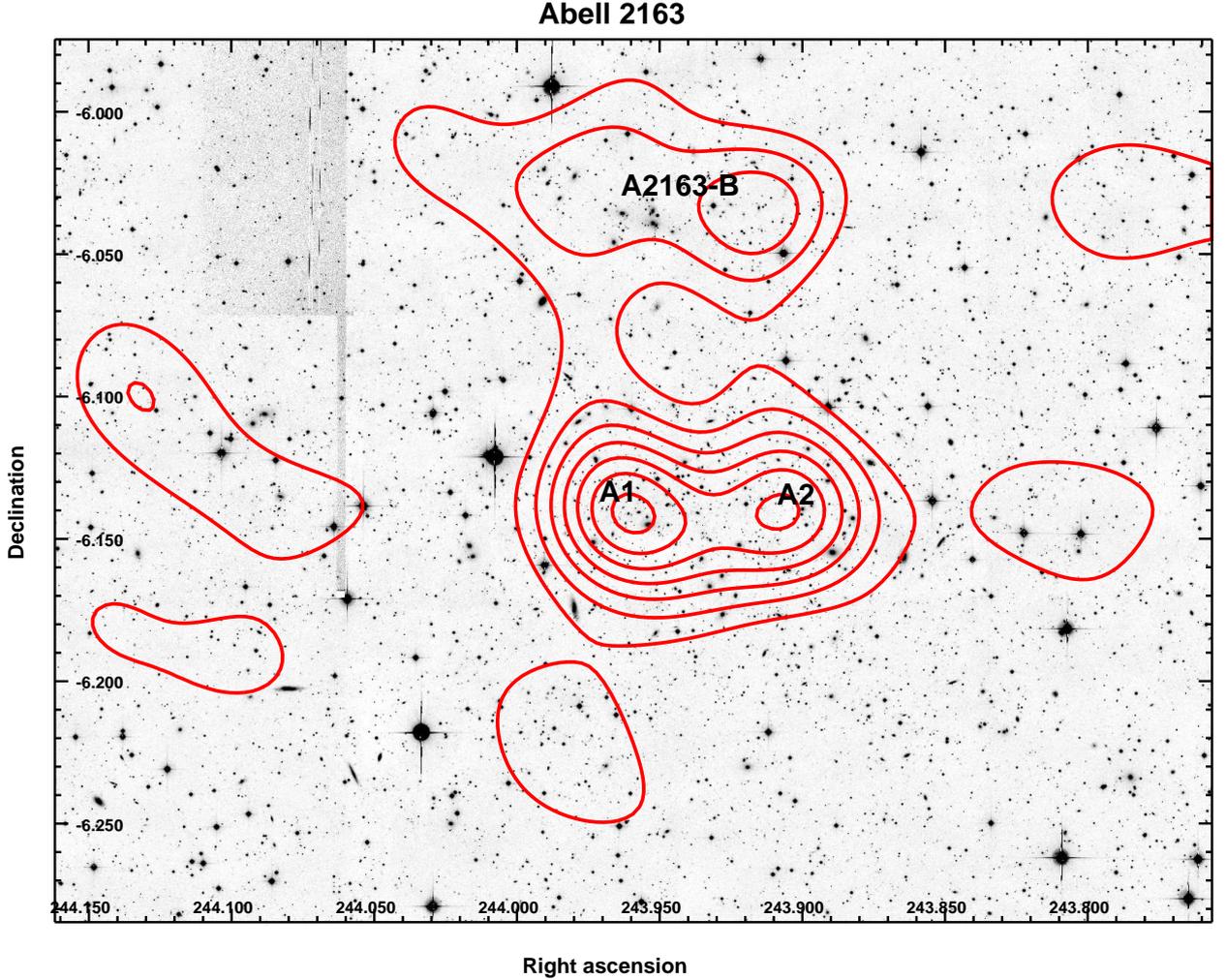}
   \caption{MegaCam r' image of the cluster Abell~2163, overlaid with 
the mass map reconstruction. The lowest contour corresponds to the 
$2\sigma$ level and the following ones are scaled with $\sigma$. The
clump A2163-B is detected at the $3.5\sigma$ level. }
  \label{fig:massmap}
\end{figure*}

To map the 2D mass distribution, we use the \textsc{LensEnt2} method,
an entropy-regularized maximum-likelihood technique developed by
\citet{marshall02}. The mass map  is built from the weak lensing catalog,
using a smoothing scale characterized by the so-called \emph{intrinsic
correlation function} (ICF) of the model. A Gaussian ICF of 160\arcsec\
($\sim 520\,h_{70}^{-1}\,\mathrm{kpc}$ at the cluster redshift) is
a good compromise between details and smoothness of the mass map,
given the number density of sources in the background catalog. The
final mass map is scaled in units of mass surface density ($M_\odot
{\mathrm pc}^{-2}$) and can be used directly to estimate the masses of
the different mass components.  The code also generates an error map that
allows the signal-to-noise ratio map to be built and the significance
of the detected peaks to be quantified.

\begin{figure}
  \centering
  \includegraphics[width=0.48\textwidth]{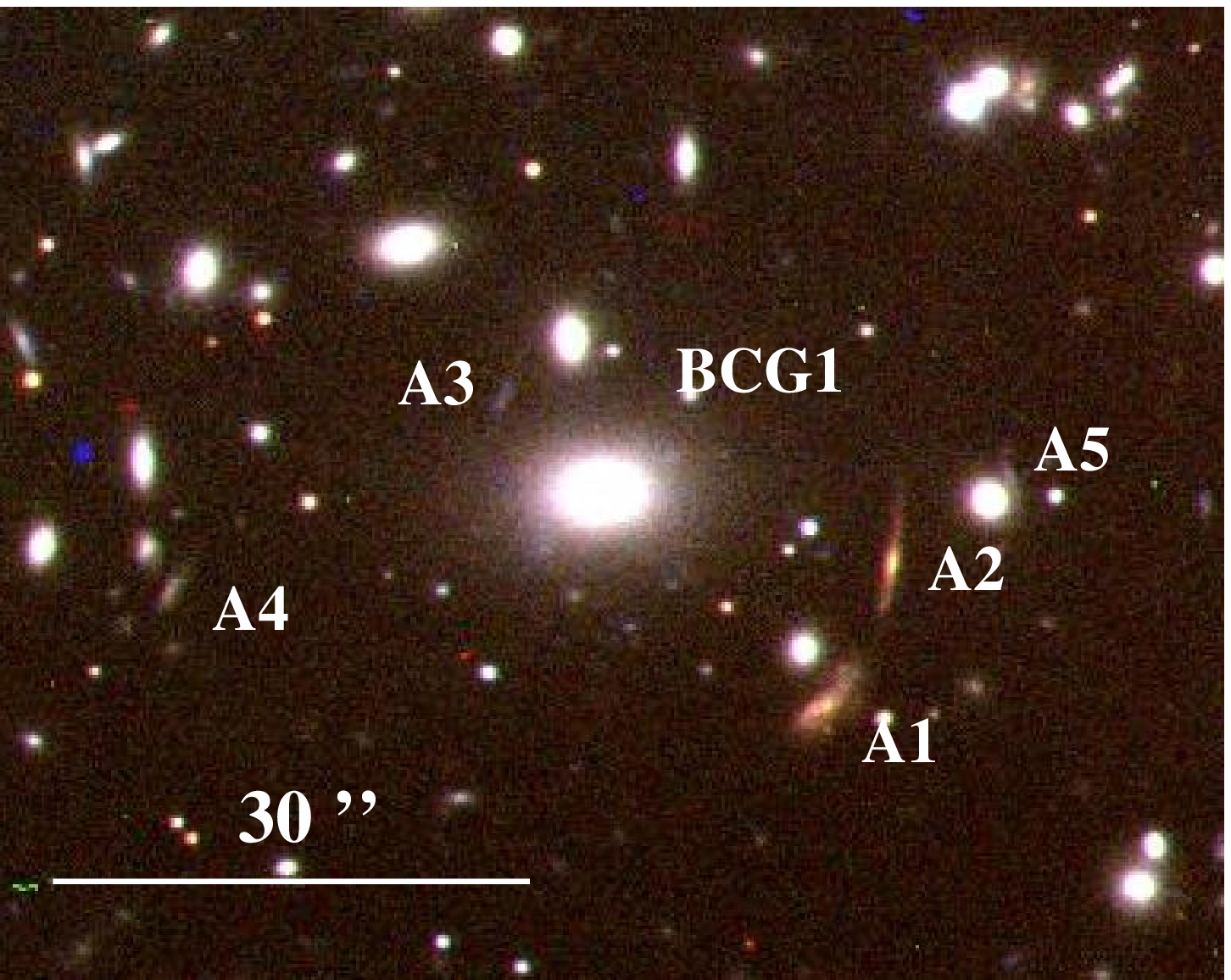}
  \caption{Combined 3-color image of the center of the cluster Abell 2163
obtained with a combination of V, R and I images taken with FORS1 at
the VLT in exceptional seeing conditions of 0.50\arcsec. In
this image, the identification of the most significant highly distorted 
features is reported. The central galaxy BCG1 is located at:
$\alpha_{2000}=16^{h}15^{m}49.0^{s}$; 
$\delta_{2000}=-06^{\circ}08{'}42.8{''}$. }
  \label{fig:vltcolor}
\end{figure}

In the field of Abell~2163, the main peak is detected at more than
$8\sigma$ and is centered close to the central galaxy BCG1. No strong
lensing features or multiple images are identified around this galaxy,
precluding any strong lensing model to explore the central mass
distribution.
 However, a few single imaged
arclets were reported  around the BCG1 \citep{fort94}, confirming that
a peak of dark matter coincides with the galaxy (Fig. \ref{fig:vltcolor}).

The mass
distribution is also significantly elongated towards the second brightest
cluster member  BCG2 (Fig.~\ref{fig:massmap}) with a second mass peak
detected at $\simeq 6 \sigma$. A detailed comparison between this mass map
and the galaxy density obtained with the galaxies selected within the ``red
sequence'' (Fig.~6 in \citet{maurogordato08}) shows a striking concordance
between the two distributions. Even the much less significant peaks in
the mass distribution are coincident with other subclusters, namely the
northern subcluster A2163-B, and other peripheral clumps C and D.
The correlation between the dark matter distribution and the cluster
galaxies density distribution clearly demonstrates that they both trace
a noncollisional component in merging clusters. These maps are very
different from the X-ray map \citep{govoni04,bourdin11}, which traces
the gas distribution with the following characteristics: a single peaked
distribution, with an E-W elongation and a peak center significantly
off-center compared to BCG1. \citet{maurogordato08} propose in their
paper a scenario in which we observe A2163 in a post-merger phase, less
than 1 Gyr after the main collision. At this stage the gas morphology
is expected to still be centrally concentrated, and elongated along the
merger axis with a possible displacement of the gas relative to the main
dark matter clump towards the secondary clump. This scenario remains fully
compatible with the present observation of the dark matter distribution
traced by weak lensing and is verified by numerical simulations of
cluster mergers including both dark matter and intracluster gas physics
\citep{roettiger97}.  Moreover, it has been recently confirmed by the
analysis of \citet{bourdin11}, who show evidence of a cool gas ``bullet"
close to the cluster center that is clearly separated from its galaxy component
and identified here as the second mass peak of the cluster.

As already mentioned, the northern secondary peak is also detected well 
in the mass map, at the $4\sigma$ level. It corresponds closely to
the clump A2163-B identified by \citet{maurogordato08} as an infalling
group with a separated velocity structure. However, there is some
uncertainty in this mass distribution, which seems strongly elongated in
this area. It is difficult to know whether this is related to an artifact
in the mass reconstruction or to a realistic mass extension. The
proximity of a bright star close to the mass clump (3\arcmin\ north
of the clump, see Fig.~\ref{fig:massmap}) reduces the available area
where background galaxies can be used for shear measurements. Moreover,
\citet{dietrich11} show that peak offsets in weak lensing maps
can occur up to 1\arcmin\ when the reconstruction is dominated by shape
noise, {\it i.e.} the statistical noise due to the intrinsic
ellipticity of the sources. This is indeed the case in this low S/N area
of the mass map. The comparison with the galaxy density distribution
(Fig.~5 in \citet{maurogordato08}) also shows similarities in this
elongated distribution,  so regardless of the exact centering of the weak
lensing mass clump, it is closely associated with the X-ray peak and the
group A2163-B infalling on the main cluster. We therefore confirm that
a significant amount of mass is associated with A2163-B. We propose to
quantify better it in the next section.

\section{Masses in Abell 2163}
\label{sec:wlmass}
\subsection{Weak lensing masses}
\begin{figure}
  \centering
  \includegraphics[width=0.35\textwidth,angle=-90]{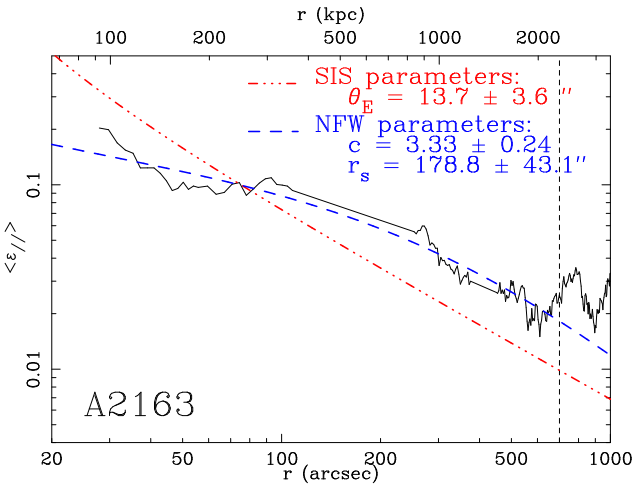}
  \caption{Measured shear profile, averaged in sliding windows
(circular annular rings of 110\arcsec\ each). Data with $100
\arcsec < r < 250\arcsec$ or $375 \arcsec < r < 450\arcsec$ have been
removed from the fit because radial signal may be perturbed by
the additional mass clumps A2 or B. The best fit with an  SIS mass 
profile and an NFW one are plotted. The fit is limited to 700\arcsec\ 
around the cluster center because the signal is not significant enough 
at a greater distance.}
  \label{fig:shearprof}
\end{figure}

There are several ways to estimate the mass from the shear
measurement of the background galaxies. The simplest one is to fit
the shear profile directly by some analytical profiles, like the one
expected from a singular isothermal sphere (SIS) or
an NFW distribution \citep{navarro97}.  These mass profiles correspond
respectively to the density profiles 
\begin{eqnarray*}
\rho_{SIS} (r) \propto \sigma^2 r^{-2} \\
\rho_{NFW} (r) = \frac{\delta_c \ \rho_c}{(r/r_s) (1 + r/r_s)^{2}} 
\end{eqnarray*}

where $\rho_c$ is the critical density at the cluster redshift. 
$\delta_c$ is related to the concentration parameter $c_{vir}$ by
\[ 
\delta_c = \frac{200}{3} \ \frac{c^3}{\ln (1+c) - \frac{c}{1+c}}
\]
and the virial radius $r_{200}$ is defined by $r_{200} = c_{vir} r_s$. 
The virial mass $M_{200}$ is the mass enclosed inside the virial
radius: 
\[M_{200} = \frac{800 \pi }{3} \ \rho_c \ r_{200}^3. \]

For all the fits, we assume spherical symmetry. However, we have
shown in the previous section that subclustering is quite significant
in the mass distribution, so we decided to remove from the fit the
data points located inside two concentric circular annuli 
corresponding to A2163-A2 ($100 \arcsec < r < 250\arcsec$) and A2163-B
($375 \arcsec < r < 450\arcsec$). We also shifted the center of the lens
1.3\arcmin\ west from BCG1 and fixed it between the two main mass peaks,
to enclose the whole mass distribution better. In practice, this shift
does not have a strong impact on the shear profile at large radius. The
shear profile is shown in Fig.~\ref{fig:shearprof}, with the
result of the best fits by the two mass profiles. In both cases, we
limited the extent of the fit to 700\arcsec\ because we consider that
the signal is not reliable at a larger distance.

\begin{figure}
  \centering
  \includegraphics[width=0.48\textwidth]{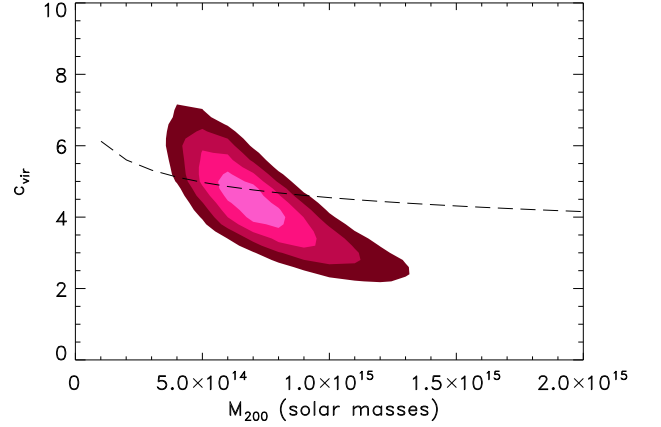}
  \caption{Distribution of the fitted parameters of the NFW profile in
the $M_{200}-c_{vir}$ plane, scaled by their likelihood. The
correlation between both parameters found from numerical simulations
of dark matter halos is plotted as a dashed line. }
  \label{fig:m200c}
\end{figure}

Instead of binning and averaging the shapes of the distorted galaxies
in circular annuli, we also implemented a global approach. We used
the \texttt{McAdam} software \citep{marshall02,marshall06}, which is
based on a Bayesian analysis of the shear distribution, approximated
by a given parametric solution. The output of \texttt{McAdam} is a
probability distribution of the fitted parameters, obtained
using a maximum-likelihood estimator and an MCMC iterative minimization.
\texttt{McAdam} works directly on the PSF-corrected faint galaxies
catalog. It takes all the galaxies into account at their position with their
shape parameters, without any ring averaging. 
Again the objects close to A2163-B are removed from the working sample,
and the mass center is shifted 1.3\arcmin\ west of BCG1, with the goal
of estimating the global component of A2163-A. 

\begin{figure}
  \centering
  \includegraphics[width=0.48\textwidth]{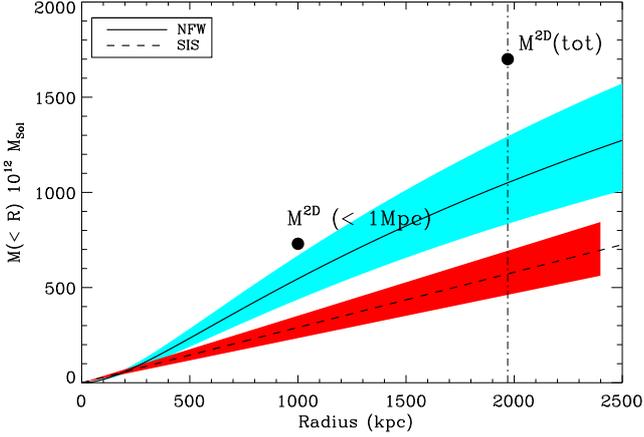}
  \caption{Integrated mass profiles obtained with
the best-fit parameters of the SIS and NFW density profiles. The values of the
projected mass directly measured on the 2D mass map and integrated up to
the radii 1 Mpc and 2 Mpc ($\simeq r_{200}$) are plotted for comparison. 
They correspond to projected masses inside cylinders so they
overestimate the mass by $\sim 25 \%$. The total mass also includes the
contribution of A2163-B with a mass of $ \sim 2.7 \
10^{14} \ h_{70}^{-1} \ M_\odot$. The vertical dashed line corresponds
to the position of the virial radius $r_{200}$.}
  \label{fig:massprof}
\end{figure}

Thanks to this global fit procedure, the shear field was fit with the SIS
and NFW distributions, assuming spherical symmetry in both cases. A prior
on the concentration parameter $c$ is included ($1.5<c<10$), following
the results of N-body simulations of cosmological structure formation
at the galaxy cluster scale \citep{bullock01,hennawi07}. Fitting the
shear profile with the SIS mass profile yields a velocity dispersion
$\sigma_\mathrm{shear}$ or, equivalently, a value for the Einstein radius
$\theta_\mathrm{E}$ scaled by the ratio $D_{\mathrm{LS}}/D_{\mathrm{OS}}$
averaged over all the sources. For the NFW mass profile, there is a
well known degeneracy between $M_{200}$, the mass enclosed within the
virial radius $r_{200}$, and the concentration parameter $c_{vir}$
(Fig.~\ref{fig:m200c}). But the average value of $c_{vir}$ resulting
from our fit ($c_{vir} = 4.6 \pm 1.6$) is consistent with the values
predicted by numerical simulations of dark matter halo properties
\citep{bullock01}.  The results of the different fits are presented in
Table~\ref{tab:fit}, and the integrated mass profiles are displayed in
Fig.\ref{fig:massprof}.

\begin{table*}
  \caption{Best-fit values obtained from the fit of the SIS and NFW
profiles. }
  \label{tab:fit} \centering 
\begin{tabular}{l|ccccc}
    \hline\hline\noalign{\smallskip}
 & & & NFW & & \\ 
 & $c_{vir}$ & $r_{200} \ ( h_{70}^{-1}$ Mpc )
 & $M_{200} \ (10^{14} \ h_{70}^{-1} \ M_\odot)$
 & $r_{500} \ ( h_{70}^{-1}$ Mpc )
 & $M_{500} \ (10^{14} \ h_{70}^{-1} \ M_\odot)$ \\
    \noalign{\smallskip}\hline\noalign{\smallskip}
Radial bins & $3.36 \pm 0.09$ & $1.97 \pm 0.2$ & $10.7 \pm 3.1$ 
& $1.27 \pm 0.12$ & $7.1 \pm 2.1$ \\
Global fit & $4.6 \pm 1.6$ & $1.79 \pm 0.23$ & $8.0 \pm 3.0$ 
& $1.18 \pm 0.13$ & $5.7 \pm 2.1$ \\
    \noalign{\smallskip}\hline\hline\noalign{\smallskip}
 & & & SIS & & \\ 
 & & $\theta_E \ (\arcsec)$ & $\sigma_{los}$ (km/s) & $M_{tot}$ \ ($r <
r_{200}$) & \\
    \noalign{\smallskip}\hline\noalign{\smallskip}
Radial bins & & $13.7 \pm 3.6$ & $810 \pm 110$ &  $600 \pm 220$ & \\
Global fit & & $13.0 \pm 2.6$ & $790 \pm 80$ & $570 \pm 180$  & \\
    \noalign{\smallskip}\hline\hline\noalign{\smallskip}
 & \multicolumn{5}{c} {$M^{\mathrm{2D}} \ (r< 1 \ \mathrm{Mpc}) \simeq
7.3 \ 10^{14} \ h_{70}^{-1} \ M_\odot$} \\
2D mass map  & \multicolumn{5}{c} {$M^{\mathrm{2D}} \
({\mathrm{A2163-A1}}) \simeq 7.1 \ 10^{14} \ h_{70}^{-1} \ M_\odot$} \\
 & \multicolumn{5}{c} {$M^{\mathrm{2D}} \ ({\mathrm{A2163-A2}}) \simeq
2.5 \ 10^{14} \ h_{70}^{-1} \ M_\odot$}  \\
 & \multicolumn{5}{c} {$M^{\mathrm{2D}} \ ({\mathrm{A2163-B}}) \simeq
2.7 \ 10^{14} \ h_{70}^{-1} \ M_\odot$} \\
 & \multicolumn{5}{c} {$M^{\mathrm{2D}} \ (r< r_{200})  \sim 17.0 \ 10^{14} \ h_{70}^{-1} \ M_\odot \ \Longleftrightarrow \ M_{\mathrm{tot}} \sim 14.0 \ 10^{14} \ h_{70}^{-1} \ M_\odot$} \\
\noalign{\smallskip}\hline
  \end{tabular}

\medskip
The first line corresponds to the 1D fit of the shear profile while the
second line gives the result of the global fit with \texttt{McAdam}.
The estimates based on the strong lensed features are indicated as a
complement. The last lines correspond to the projected masses measured
directly on the 2D mass map. They correspond to masses measured inside
a cylinder of radius 1 Mpc and 2.0 Mpc ($=r_{200}$), respectively, as
well as the projected mass of the different mass components A2163-A1,
A2163-A2 and A2163-B.  
\end{table*}

The  masses of the different mass clumps identified in the 2D mass map
(Fig.~\ref{fig:massmap}) are also measured through direct integration
of the mass distribution. No special shape is assumed and the mass is
integrated in hand-defined contours or within given apertures. In a first
attempt, we integrated the mass of the main clump in concentric annuli up
to 300\arcsec ($\simeq 1$ Mpc) and centered between the two mass peaks
A1 and A2. This radius corresponds to the limit where A2163-B starts
to be detected in the map, and it includes both mass peaks (A1 and A2).
We find a projected mass of $7.3 \ 10^{14} \ h_{70}^{-1} \ M_\odot$, which
is the mass integrated within a cylinder of radius 1 Mpc.  This projected
mass exceeds the mass integrated inside a sphere of the same radius
by approximately 25\%\ for a standard NFW profile.  In comparison,
the projected mass deduced from the previously fitted values of the
NFW profile and enclosed in a 1 Mpc radius is $8.0 \pm 2.3 \ 10^{14}
\ h_{70}^{-1} \ M_\odot$, consistent with the direct value.  We also
measured the total mass inside a cylinder defined by the ``virial radius"
$r_{200}$ taken as 2.0 Mpc.  This mass $\sim 1.7 \ 10^{15} \ h_{70}^{-1}
\ M_\odot$ includes the contributions of the two mass clumps, A2163-A
(A1 and A2) and A2163-B.  It reduces to  $M_{tot} \sim 1.4 \ 10^{15} \
h_{70}^{-1} \ M_\odot$ for the 3D mass inside a sphere of radius $r_{200}$
centered between A1 and A2. This last value is considered as the total
mass of the cluster within the virial radius (Table~\ref{tab:fit}).

Going further in the mass measures from the 2D mass map, we also attempted
to estimate separately the masses of each of the two mass subclumps
that we have identified in A2163-A (A1 and A2).  These are approximate
measures because they are spatially limited and do not properly include
the large-scale extension of the cluster.  The measured masses are $7.1
\ 10^{14} \ h_{70}^{-1} \ M_\odot$ and $2.5 \ 10^{14} \ h_{70}^{-1}
\ M_\odot$ respectively, and their sum represents a major fraction
of the total mass. Interestingly, we are able to measure a mass ratio
$\sim$ 3:1 between the 2 clumps, although this result must be taken with
caution. This is not too different from the mass ratio 4:1 estimated by
\citet{bourdin11} from interpretating of the merging scenario. Our ratio
3:1 also seems closer to the apparent ratio in the galaxy distribution
and could be more representative of the total mass repartition.

For the clump A2163-B, we integrated the mass within a hand-designed
contour close to a circular annulus of 500 kpc, in order to avoid the
contamination by the main cluster. This gives a mass of $\sim 2.7 \
10^{14} \ h_{70}^{-1} \ M_\odot$, so slightly higher than the
mass estimated by \citet{bourdin11}. However this is a rather uncertain
measure, since the weak lensing signal and its distribution are not fully
reliable.

All these results are summarized in Table~\ref{tab:fit}. It is interesting
to note that the sum of the direct mass of the individual clumps (A1 +
A2 + B) represents $\sim$ 75\% of the total mass. The remaining mass
represents the large-scale extension of the distribution and possibly
the additional clumps contribution. Moreover, although there are large
uncertainties in the direct approach, the results are consistent with
those derived from the fit with usual mass distributions.

\subsection{Comparison with other lensing mass measurements}
It has long been pointed that, regardless of the method used to determine
the mass of A2163, the lensing estimates are much lower than the values
expected for this very hot cluster. For example, \citet{squires97} fitted
their data by an SIS profile with a velocity dispersion $\sigma_{SIS} =
740$ km/s, and \citet{cypriano04} found $\sigma = 1020 \pm 150$ km/s. Both
values are compatible with the present measure, even if they were obtained
with fewer galaxies: 261 background galaxies for \citet{cypriano04} and
736 galaxies for \citet{squires97}. We are more skeptical of the fitted
values proposed by \citet{radovich08} derived from the same initial data
as in the present work. With our selection process and the great care
we took for the PSF correction, we ended with a higher galaxy density
(12 instead of 8 galaxies/arcmin$^2$), providing the largest sample
of background sources used in the weak lensing mass determination in
A2163 14550 galaxies within 0.38 square degree).  Moreover, looking into
details in the shear profile in \citet{radovich08}, we note that their
fit starts with data points located at a much larger radial distance
than ours. From our study, removing data points below $r<200\arcsec$ in
the fit of the SIS profile increases the value of $\sigma$ to 1100 km/s,
close to the value found by \citet{radovich08}. The quality of the fit is
very poor so we consider that the large-scale mass distribution in A2163
is too complex and shallow to be fitted by a single SIS profile. More
recently, \citet{okabe11} have also performed a weak lensing analysis
using Subaru imaging data. They confirm the bimodal structure of the mass
in the central parts of the cluster, although the mass ratio between A1
and A2 is much higher (8:1 to 10:1), and the detection of the clump A2
is not as significant as in our own map. The 2D mass extractions from
weak lensing reconstructions are still limited in their quantitative use
and would strongly benefit from the higher source density of HST imaging
\citep{becker11}.  The total mass estimates obtained by \citet{okabe11}
are also higher than the present ones, at least by 50\%\ when they use
a multicomponent analysis and by a factor 2 for a single mass profile
reconstruction. But in the last case, their fit show high mass associated
with a low value of the concentration parameter ($c \sim 2.8$). This
combination of fitted parameters is probably representative of the
mass-concentration degeneracy shown in Fig~\ref{fig:m200c}.

In conclusion, there are still some discrepancies between different
approaches used to infer quantitatively the total mass of this
complex cluster, but the more recent results become more consistent
with each other, provided similar criteria are taken into account to
define a ``total mass''. Absolute values for the mass still suffer from
uncertainties in the large-scale structures' contributions and are also
highly dependent on the intrinsic noise owing to the limited number of
background sources: \citet{hoekstra11} show that it can reach a 20 \%\
to 25 \%\ level for ground-based observations, whatever the methodology
established to produce the results.

\subsection{Galaxies, gas and dark matter distributions: consequences
on the merger scenario} 
\begin{figure*}
  \includegraphics[width=18cm]{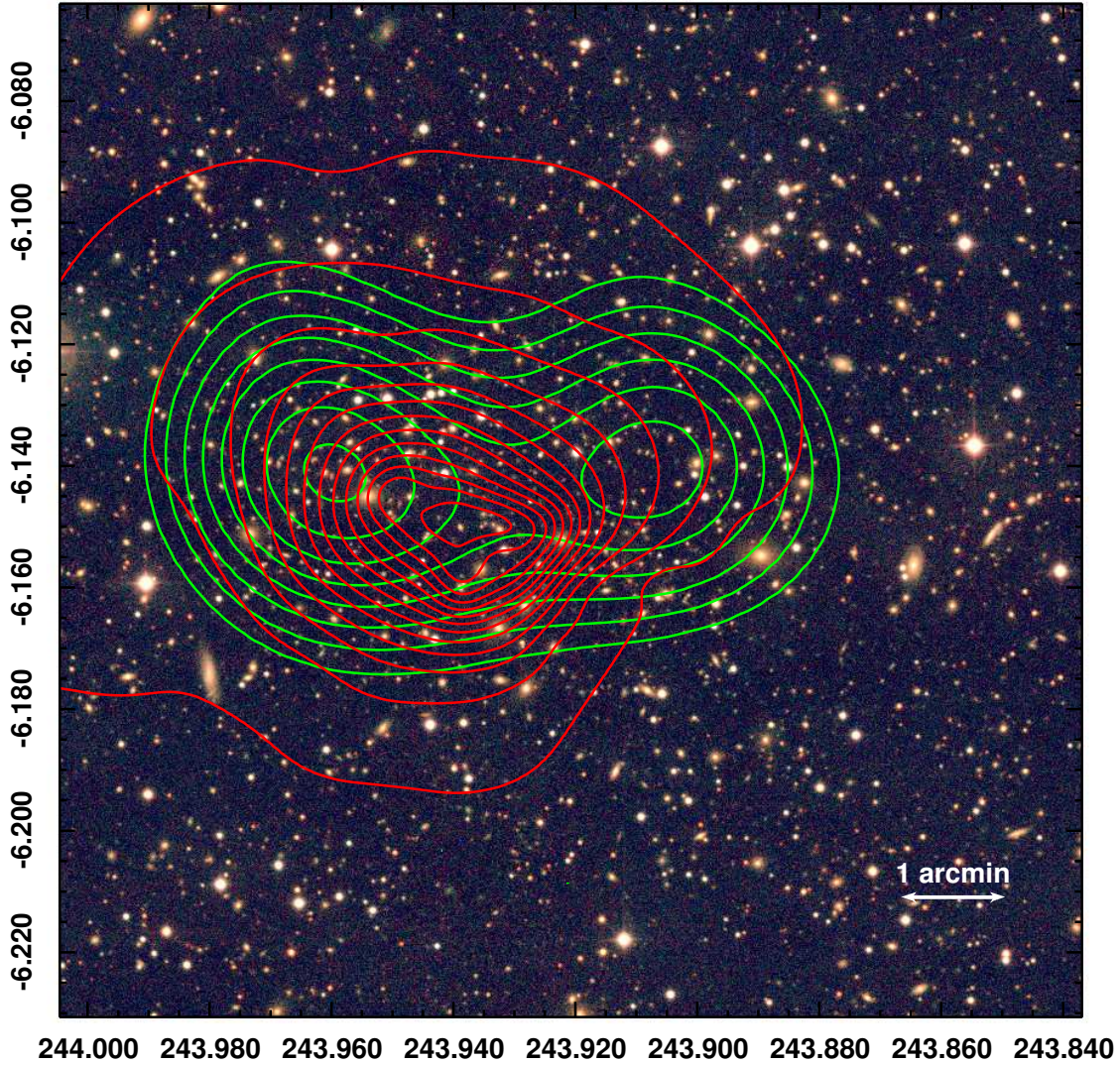}
  \caption{Multi-color image of the center of the cluster Abell 2163
with overlays of the dark matter distribution (green contours from the
lensing map) and of
the gas distribution (red contours for the Chandra X-rays image,
courtesy H. Bourdin). The shift between the two mass clumps is
3\arcmin , while the distance between the X-ray maximum and the eastern
clump A2163-A1 is 1.2\arcmin . The total size is 10\arcmin x
10\arcmin. 
}
  \label{fig:all_dist}
\end{figure*}

We focus in this section on a comparison between the masses derived from
weak lensing, which trace the dark matter in the cluster, the masses
derived from X-rays, and the spatial distribution of the galaxies. The
X-ray temperature  is very high ($T_X \sim 12$ to 14 keV) regardless of the
instruments used to measure it \citep{elbaz95,govoni04}. The temperature
distribution shows strong variations in the central part of the cluster
\citep{markevitch96a,markevitch01} and a significant nonthermal component
at high energy \citep{rephaeli06}. Extrapolating the mass under these
conditions is a risky exercise and must be taken with caution because
the cluster is obviously not in thermal equilibrium. To overcome this
difficulty, \citet{bourdin11} have computed the Y$_X$ parameter defined by
\citet{kravtsov06}. Y$_X$ is considered as one of the best proxies
for the total cluster mass. They used different calibrations of the
scaling relation $Y_X - M_{500}$ \citep{arnaud10, vikhlinin09} to give a
total mass estimate that does not rely explicitly on the hydrostatic
equilibrium hypothesis. Regardless of the calibration used, they find a
value $M_{500} \simeq 1.8$ to $2.0 \ 10^{15}$ M$_\odot$, {\it i.e.}.
This converts to a mass $M_{200}$ within the virial radius 50\%\
higher for a typical NFW mass profile, so this mass is two to three 
times higher than the lensing value. It is not
clear why there is still such a significant discrepancy between
the weak lensing mass estimates and the ones derived from X-ray proxies.

The spatial distributions of the three main components of the
cluster are displayed in Fig~\ref{fig:all_dist}: the intracluster gas
\citep{bourdin11} and the dark matter (this paper) distributions are
overlaid on the galaxy cluster image. As already observed in several
other merging clusters, these distributions show evidence that the
baryonic mass (the gas) does not follow the total mass (dominated by the
dark matter). On the contrary, the galaxies distribution is spatially
more coherent with the dark matter distribution A bi-dimensional
Kolmogorov-Smirnov test \citep{fasano87} could be applied to get
a more quantitative answer but his needs further adjustments in
the galaxies selection, which are beyond the scope of this paper.
\citet{maurogordato08} show morphological differences in the galaxy
density distribution for different luminosity bins that require deeper
insights to reconcile with the dark matter distribution.  Nevertheless, in
contrast to the intracluster gas, both galaxies and dark matter represent
noncollisional mass components so we confirm that the total mass of the
cluster is dominated by a collisionless form of dark matter.  This is at
least the fourth  example of a cluster merger with such properties, after
the initial work on the ``Bullet cluster'' 1E 0657--56 \citep{bradac06}
followed by similar trends observed in Abell 520 \citep{mahdavi07}
and MACS J0025.4--1222 \citep{bradac08}.

Abell 2163 is the cluster with the strongest discrepancy between
X-rays and lensing masses. It is hard to reconcile both values, even when
knowing the different limitations discussed above, and we still do not
have a clear vision of the physical origin of the discrepancy. Because of
the merging process, the gas distribution could be far from isothermal on
large scales, and this might induce a strong uncertainty, which depends on
the exact phase of the cluster merger \citep{poole07}.  We suspect that
all absolute mass determinations with X-ray proxies in a strong merger
like A2163 must be taken with caution \citep{okabe10}, and the gas mass
must be revisited in the framework of a colliding cluster. We also
expect better lensing mass estimates using deep HST data. However,
apart from a difference in the absolute mass scaling, the relative
distributions of the three components are consistent with each other
in a general scenario of a cluster major merger.

\section{Conclusion}
\label{sec:conclusions}
We have presented in this paper a detailed dark matter map of
the cluster Abell 2163, using the weak lensing effect on background
sources. We provide global masses, as well as the bi-dimensional mass
distribution and mass estimates for each individual substructure
identified in the map:
\begin{itemize}
\item Thanks to a detailed mass reconstruction, we were able to separate
the central mass clump into two components with a distribution in agreement
with the cluster galaxies density distribution \citep{maurogordato08}. The
mass of the main clump is about $1.0 \ 10^{15} \ h_{70}^{-1} \ M_\odot$
and is split in two clumps with a mass ration $\sim$ 3:1. It is
consistent with previous weak lensing estimates, but does not follow the
expected correlations with the X-ray properties of the gas. Such a mass
discrepancy is a strong argument in favor of the merging scenario in
the cluster.  The second mass clump A2163-A2 is shifted compared to the
cold gas core detected by \citet{bourdin11}, and this shift can be due
to the trailing of the gas compared to the noncollisional dark matter
during the crossing of the main clump.

\item We detect at more than 4$\sigma$ the dark matter associated
with the clump 2163-B, identified spectroscopically by
\citet{maurogordato08}. This clump has a mass value
($\simeq 2.7 \ 10^{14} \ h_{70}^{-1} \ M_\odot$) compatible with a small
cluster. This is also coherent with the X-ray temperature on the order
of 4 keV identified by \citet{bourdin11}.  As already suggested by
\citet{maurogordato08}, this clump has not yet undergone its main
interaction with the central cluster and is still infalling towards
the center.

\item Concerning the total mass of the whole complex, our mass estimates
do not exceed  $1.5 \ 10^{15} \ h_{70}^{-1} \ M_\odot$. This value is
well below the expected mass given by the correlations with the X-ray
properties of the cluster. We confirm that the scaling laws that use
X-ray proxies to infer the mass of clusters must be taken with caution
for structures that are far from the hydrostatic equilibrium. 
\end{itemize}

The results of our study are a new step forward in the process of building
a coherent scenario for the multiple mergers in the cluster Abell
2163. This thorough analysis of the weak lensing mass reconstruction
was possible thanks to the combination of both the spatial extent and
the depth of the data.  It confirms the bimodal structure of the matter
distribution in the cluster center and the physical separation between
the baryonic mass and the noncollisional mass components. Our results
also show that there exists many similarities between the gas and
dark matter distributions in Abell~2163 and in the ``Bullet cluster''
1E0657--56.  Both cases display a gas core spatially separated from its
dark matter component, even if the mass ratio between the clumps, the 3D
configuration of the merger and the age of the merger differ. But they
offer different merging configurations that can feed the comparisons with
numerical simulations \citep{poole07}. Moreover, Abell~2163 presents
an additional complexity with the identification of several other
mass clumps involved in the merger but at an earlier stage or in the
premerging phase. In conclusion, Abell~2163 has long been considered
as an exceptional cluster because of its high gas temperature. It is
now more interesting to explore the richness and the complexity of the
physical processes occurring within it.

\begin{acknowledgements}
I wish to thank Phil Marshall for his help in implementing the code {\sc
Lensent2} and Monique Arnaud, Marceau Limousin, Sophie Maurogordato,
Roser Pell\'o, Etienne Pointecouteau, and Gabriel Pratt for fruitful
discussions. Herv\'e Bourdin kindly shared his X-ray data with us, and we
benefited from his extensive knowledge. I also thank an anonymous referee
for his careful reading of the paper and his pertinent questions. Finally
I thank the Programme National de Cosmologie of the CNRS for financial
support.  \end{acknowledgements}

\end{document}